\pdfoutput=1
\documentclass[12pt,a4paper]{article}

\usepackage{ifthen} 
\usepackage{graphicx}
\usepackage{subfig}
\newboolean{pdflatex}
\setboolean{pdflatex}{true} 

\newboolean{articletitles}
\setboolean{articletitles}{true} 

\newboolean{uprightparticles}
\setboolean{uprightparticles}{false} 

\def\paperauthors{LHCb collaboration} 
\def\paperasciititle{Metal Foil Detectors assembly for the beam and background monitoring in the LHCb experiment} 
\def\papertitle{Metal Foil Detectors assembly for the beam and background monitoring in the \lhcb experiment} 
\def\paperkeywords{
{High Energy Physics},
{LHCb},
{Hardware and accelerator control systems},
{Radiation-hard detectors},
{Radiation-induced secondary-electron emission},
{Beam-line instrumentation (beam position and profile monitors, beam-intensity monitors)}
} 
\def\papercopyright{\the\year\ CERN for the benefit of the LHCb collaboration} 
\def\paperlicence{CC BY 4.0 licence}
\def\paperlicenceurl{https://creativecommons.org/licenses/by/4.0/}

\newif\ifEnableSectionTOCLinks
\EnableSectionTOCLinksfalse 

\usepackage[top=1in, bottom=1.25in, left=1in, right=1in]{geometry}

\usepackage{microtype}
\usepackage{lineno}  
\usepackage{xspace} 
\usepackage{caption} 

\usepackage{graphicx}  
\usepackage{color}
\usepackage{colortbl}
\graphicspath{{./figs/}} 

\usepackage{amsmath} 
\usepackage{amssymb}
\usepackage{amsfonts}
\usepackage{upgreek} 

\newcommand*\patchAmsMathEnvironmentForLineno[1]{%
\expandafter\let\csname old#1\expandafter\endcsname\csname #1\endcsname
\expandafter\let\csname oldend#1\expandafter\endcsname\csname
end#1\endcsname
 \renewenvironment{#1}%
   {\linenomath\csname old#1\endcsname}%
   {\csname oldend#1\endcsname\endlinenomath}%
}
\newcommand*\patchBothAmsMathEnvironmentsForLineno[1]{%
  \patchAmsMathEnvironmentForLineno{#1}%
  \patchAmsMathEnvironmentForLineno{#1*}%
}
\AtBeginDocument{%
\patchBothAmsMathEnvironmentsForLineno{equation}%
\patchBothAmsMathEnvironmentsForLineno{align}%
\patchBothAmsMathEnvironmentsForLineno{flalign}%
\patchBothAmsMathEnvironmentsForLineno{alignat}%
\patchBothAmsMathEnvironmentsForLineno{gather}%
\patchBothAmsMathEnvironmentsForLineno{multline}%
\patchBothAmsMathEnvironmentsForLineno{eqnarray}%
}

\usepackage[pdftex,
            pdfauthor={\paperauthors},
            pdftitle={\paperasciititle},
            pdfkeywords={\paperkeywords}]{hyperref}
\usepackage{hyperxmp}
\hypersetup{
    pdfcopyright={Copyright (C) \papercopyright},
    pdflicenseurl={\paperlicenceurl}
}

\usepackage[bottom,flushmargin,hang,multiple]{footmisc}

\usepackage[all]{hypcap} 

\usepackage{xspace} 
\usepackage{upgreek}
\def\lhcb   {\mbox{LHCb}\xspace}
\def\atlas  {\mbox{ATLAS}\xspace}
\def\cms    {\mbox{CMS}\xspace}

\def\cern {\mbox{CERN}\xspace}
\def\lhc    {\mbox{LHC}\xspace}

\def\runone {\mbox{Run~1}\xspace}
\def\runtwo {\mbox{Run~2}\xspace}
\def\runthree {\mbox{Run~3}\xspace}

\def\upgradeone {\mbox{Upgrade~I}\xspace}

\def\plume  {PLUME\xspace}
\def\rmsthree {RMS-R3\xspace}
\def\MagUp {\mbox{\em Mag\kern -0.05em Up}\xspace}

\def\ecs    {ECS\xspace}

\ifthenelse{\boolean{uprightparticles}}%
{

 \def\PDelta      {\ensuremath{\Delta}\xspace}                 
 \def\PXi         {\ensuremath{\Xi}\xspace}                 
 \def\PLambda     {\ensuremath{\Lambda}\xspace}                 
 \def\PSigma      {\ensuremath{\Sigma}\xspace}                 
 \def\POmega      {\ensuremath{\Omega}\xspace}                 
 \def\PUpsilon    {\ensuremath{\Upsilon}\xspace}
 \let\oldPi\Pi
 \def\PPi         {\ensuremath{\oldPi}\xspace}                 

 \def\PB      {\ensuremath{\mathrm{B}}\xspace}                 
 \def\PD      {\ensuremath{\mathrm{D}}\xspace}                 

 \def\PK      {\ensuremath{\mathrm{K}}\xspace}                 
 \def\Ps      {\ensuremath{\mathrm{s}}\xspace}

 \def\thebaroffset{0.0em}
}
{

 \mathchardef\PDelta="7101
 \mathchardef\PXi="7104
 \mathchardef\PLambda="7103
 \mathchardef\PSigma="7106
 \mathchardef\POmega="710A
 \mathchardef\PUpsilon="7107
 \mathchardef\PPi="7105
 \def\PB      {\ensuremath{B}\xspace}                 
 \def\PD      {\ensuremath{D}\xspace}                 

 \def\PK      {\ensuremath{K}\xspace}                 
 \def\Ps      {\ensuremath{s}\xspace}

 \def\thebaroffset{0.18em}
}
\newcommand{\offsetoverline}[2][\thebaroffset]{\kern #1\overline{\kern -#1 #2}}%

\makeatletter
\ifcase \@ptsize \relax
  \newcommand{\miniscule}{\@setfontsize\miniscule{4}{5}}
\or
  \newcommand{\miniscule}{\@setfontsize\miniscule{5}{6}}
\or
  \newcommand{\miniscule}{\@setfontsize\miniscule{5}{6}}
\fi
\makeatother

\DeclareRobustCommand{\optbar}[1]{\shortstack{{\miniscule (\rule[.5ex]{1.25em}{.18mm})}
  \\ [-.7ex] $#1$}}

\def\squark    {{\ensuremath{\Ps}}\xspace}

\def\KorKbar {\kern \thebaroffset\optbar{\kern -\thebaroffset \PK}{}\xspace}

\def\D       {{\ensuremath{\PD}}\xspace}

\def\DorDbar {\kern \thebaroffset\optbar{\kern -\thebaroffset \PD}\xspace}

\def\Dp      {{\ensuremath{\D^+}}\xspace}
\def\Dm      {{\ensuremath{\D^-}}\xspace}

\def\DpDm    {\ensuremath{\Dp {\kern -0.16em \Dm}}\xspace}

\def\B       {{\ensuremath{\PB}}\xspace}

\def\BorBbar {\kern \thebaroffset\optbar{\kern -\thebaroffset \PB}\xspace}

\def\Bd      {{\ensuremath{\B^0}}\xspace}

\def\BdorBdbar {\kern \thebaroffset\optbar{\kern -\thebaroffset \Bd}\xspace}

\def\Bs      {{\ensuremath{\B^0_\squark}}\xspace}

\def\BsorBsbar {\kern \thebaroffset\optbar{\kern -\thebaroffset \Bs}\xspace}

\def\Y#1S{\ensuremath{\PUpsilon{(#1S)}}\xspace}

\def\LorLbar     {\kern \thebaroffset\optbar{\kern -\thebaroffset \PLambda}\xspace}

\def\CP                {{\ensuremath{C\!P}}\xspace}

\def\AT#1     {\ensuremath{A_{\mathrm{T}}^{#1}}\xspace}           

\def\C#1      {\ensuremath{\mathcal{C}_{#1}}\xspace}                       
\def\Cp#1     {\ensuremath{\mathcal{C}_{#1}^{'}}\xspace}                    
\def\Ceff#1   {\ensuremath{\mathcal{C}_{#1}^{\mathrm{(eff)}}}\xspace}        
\def\Cpeff#1  {\ensuremath{\mathcal{C}_{#1}^{'\mathrm{(eff)}}}\xspace}       
\def\Ope#1    {\ensuremath{\mathcal{O}_{#1}}\xspace}                       
\def\Opep#1   {\ensuremath{\mathcal{O}_{#1}^{'}}\xspace}                    

\newcommand{\nospaceunit}[1]{\ensuremath{\text{#1}}}       
\newcommand{\aunit}[1]{\ensuremath{\text{\,#1}}}       

\newcommand{\tev}{\aunit{Te\kern -0.1em V}\xspace}

\newcommand{\gev}{\aunit{Ge\kern -0.1em V}\xspace}
\newcommand{\mev}{\aunit{Me\kern -0.1em V}\xspace}
\newcommand{\kev}{\aunit{ke\kern -0.1em V}\xspace}
\newcommand{\ev}{\aunit{e\kern -0.1em V}\xspace}
 
\newcommand{\mevc}{\ensuremath{\aunit{Me\kern -0.1em V\!/}c}\xspace}
\newcommand{\gevc}{\ensuremath{\aunit{Ge\kern -0.1em V\!/}c}\xspace}
\newcommand{\mevcc}{\ensuremath{\aunit{Me\kern -0.1em V\!/}c^2}\xspace}
\newcommand{\gevcc}{\ensuremath{\aunit{Ge\kern -0.1em V\!/}c^2}\xspace}

\def\m    {\aunit{m}\xspace}

\def\cma  {\ensuremath{\aunit{cm}^2}\xspace}
\def\mm   {\aunit{mm}\xspace}

\def\mum  {\ensuremath{\,\upmu\nospaceunit{m}}\xspace}

\def\mus  {\ensuremath{\,\upmu\nospaceunit{s}}\xspace}

\def\mhz  {\ensuremath{\aunit{MHz}}\xspace}
\def\khz  {\ensuremath{\aunit{kHz}}\xspace}
\def\hz   {\ensuremath{\aunit{Hz}}\xspace}

\def\Xrad {\ensuremath{X_0}\xspace}

\def\mip {MIP\xspace}

\def\deriv {\ensuremath{\mathrm{d}}}

\def\gsim{{~\raise.15em\hbox{$>$}\kern-.85em
          \lower.35em\hbox{$\sim$}~}\xspace}
\def\lsim{{~\raise.15em\hbox{$<$}\kern-.85em
          \lower.35em\hbox{$\sim$}~}\xspace}

\newcommand{\lum} {\ensuremath{\mathcal{L}}\xspace}
\def\instlumi  {\ensuremath{\aunit{cm}^{-2}\!\aunit{s}^{-1}}\xspace}

\def\tell1  {TELL1\xspace}
\def\ukl1   {UKL1\xspace}

\newcommand{\etc}{\mbox{\itshape etc.}\xspace}

\newcommand{\lhcborcid}[1]{\href{https://orcid.org/#1}{\hspace*{0.1em}\raisebox{-0.45ex}{\includegraphics[width=1em]{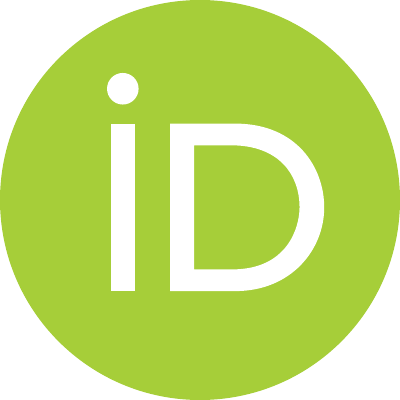}}}}

\hypersetup{
  colorlinks   = true, 
  urlcolor     = blue, 
  linkcolor    = blue, 
  citecolor    = red   
}

\ifEnableSectionTOCLinks
    \usepackage[explicit]{titlesec} 
    
    \let\oldcontentsline\contentsline
    \renewcommand

    \titleformat{\section}{\normalfont\Large\bf}{\hyperlink{tocsection.\thesection}{{\thesection} \parbox[t]{\dimexpr\textwidth-1pc}{#1}}}{1pc}{}

    \titleformat{\subsection}{\normalfont\bf}{\hyperlink{tocsubsection.\thesubsection}{{\thesubsection} \parbox[t]{\dimexpr\textwidth-1pc}{#1}}}{1pc}{}

    \titleformat{name=\section,numberless}[display]{}{}{0pt}{\normalfont\Huge\bfseries #1}
\fi

\usepackage{cite} 
\usepackage{mciteplus}

\begin{document}

\renewcommand{\thefootnote}{\fnsymbol{footnote}}
\setcounter{footnote}{1}


\begin{titlepage}
\pagenumbering{roman}

\vspace*{-1.5cm}
\centerline{\large EUROPEAN ORGANIZATION FOR NUCLEAR RESEARCH (CERN)}
\vspace*{1.5cm}
\noindent
\begin{tabular*}{\linewidth}{lc@{\extracolsep{\fill}}r@{\extracolsep{0pt}}}
\ifthenelse{\boolean{pdflatex}}
{\vspace*{-1.5cm}\mbox{\!\!\!\includegraphics[width=.14\textwidth]{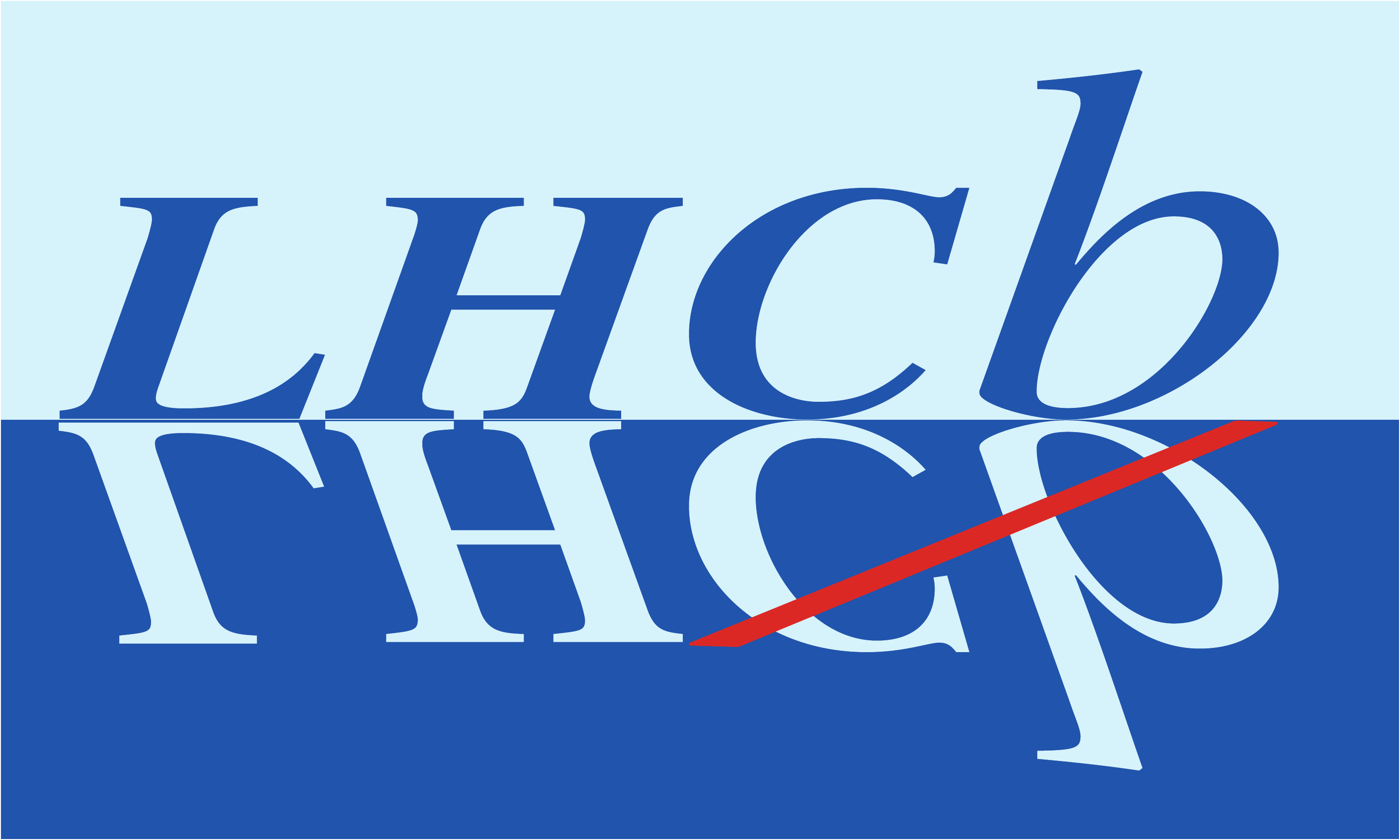}} & &}%
{\vspace*{-1.2cm}\mbox{\!\!\!\includegraphics[width=.12\textwidth]{lhcb-logo.pdf}} & &}%
\\
 & & CERN-LHCb-DP-2025-005 \\  
 & & 11 July 2025\\ 
 & & \\
\end{tabular*}

\vspace*{2.0cm}

{\normalfont\bfseries\boldmath\huge
\begin{center}
  \papertitle 
\end{center}
}

\vspace*{1.0cm}

\begin{center}
\begin{flushleft}
V.~Pugatch$^{a,*}$\lhcborcid{0000-0002-5204-9821},
F.~Alessio$^{b}$\lhcborcid{0000-0001-5317-1098},
V.~Balagura$^{c}$\lhcborcid{0000-0002-1611-7188},
F.~Blanc$^{d}$\lhcborcid{0000-0001-5775-3132},
S.~Chernyshenko$^{a}$\lhcborcid{0000-0002-2546-6080},
V.~Dobishuk$^{a}$\lhcborcid{0000-0001-9004-3255},
V.~Kyva$^{a}$\lhcborcid{0009-0006-5314-6419},
O.~Okhrimenko$^{a}$\lhcborcid{0000-0002-0657-6962},
D.~Ramazanov$^{a}$\lhcborcid{0000-0002-0497-4860},
H.~Schindler$^{b}$\lhcborcid{0000-0002-1468-0479},
O.~Schneide$^{d}$\lhcborcid{0000-0002-6014-7552}
\bigskip

{\footnotesize \it
$^{a}$HEP Department, Institute for Nuclear Research of the NAS of Ukraine,\\
\phantom{$^{a}$}47, Prospekt Nauky, 03028 Kyiv, Ukraine\\
$^{b}$EP Department, European Organization for Nuclear Research (CERN),\\
\phantom{$^{b}$}Gen\`eve 23, 1211 Geneva, Switzerland\\
$^{c}$Laboratoire Leprince-Ringuet, CNRS/IN2P3, Ecole Polytechnique, Institut Polytechnique de Paris,\\
\phantom{$^{c}$}Av. Chasles, 91120 Palaiseau, France\\
$^{d}$High Energy Physics Laboratory, Ecole Polytechnique F\'{e}d\'{e}rale de Lausanne (EPFL),\\
\phantom{$^{d}$}BSP - Cubotron, 1015 Lausanne, Switzerland\\
\medskip
$^{*}$Corresponding author:
\href{mailto:pugatch@kinr.kiev.ua}{pugatch@kinr.kiev.ua}
}
\end{flushleft}

\end{center}

\vspace*{0.5cm}

\begin{abstract}
\noindent
After an upgrade in 2019--2021, the \lhcb experiment is taking data in \runthree (2022--2026) with an instantaneous luminosity of proton-proton collisions of \mbox{$2\!\times\!10^{33}\instlumi$}.
This article presents the Radiation Monitoring System (\rmsthree) for controlling the beam and background conditions at \lhcb.
It runs continuously during the detector's operation, and independently of the main \lhcb data acquisition.
Its design is based on robust and radiation-hard Metal Foil Detector technology.
The \rmsthree monitors the instantaneous luminosity and its evolution.
The analysis of the \rmsthree \runthree data demonstrates its linear response with a high reproducibility in a five orders of magnitude dynamic range of luminosity over a long period of operation.

\end{abstract}

\vspace*{1.0cm}

\begin{center}
\small
  Published in JINST\\
  DOI \href{https://doi.org/10.1088/1748-0221/20/07/P07027}{10.1088/1748-0221/20/07/P07027}
\end{center}

\vspace{\fill}

{\footnotesize 
\centerline{\copyright~\papercopyright. \href{\paperlicenceurl}{\paperlicence}.}}
\vspace*{2mm}

\end{titlepage}


%
%
%
%


\renewcommand{\thefootnote}{\arabic{footnote}}
\setcounter{footnote}{0}

\tableofcontents



\pagestyle{plain} 
\setcounter{page}{1}
\pagenumbering{arabic}


\section{Introduction}
\label{sec:intro}
The \lhcb experiment is primarily aimed at accurately measuring \CP violation in the decays of particles with heavy quarks.
Their unprecedented statistics accumulated at \lhcb during \runone (2010--2012) and \runtwo (2015--2018) allowed to perform many important measurements related to the Cabibbo-Kobayashi-Maskawa matrix, responsible for \CP violation in the Standard Model.
\lhcb also plays a significant and unique role in the studies of the lepton flavour universality, rare decays and the searches for processes beyond the Standard Model~\cite{LHCb-PAPER-2012-031, LHCb:2022ine}.

The masses of charm and beauty quarks are relatively low compared to the \lhc energies, and have low transverse momenta.
The studied particles are most often produced close to the beams with a lot of background.
Therefore, the \lhc sets the number of interactions per bunch-crossing at \lhcb much lower than in the \atlas and \cms detectors, which target heavy particles like Higgs bosons.
However, upgrading \lhcb in 2019--2021 to cutting-edge technologies in triggering and event reconstruction allowed an increase in the number of interactions per bunch crossing from about one to five.
After the commissioning period, the target luminosity of \mbox{$2\!\times\!10^{33}\instlumi$} has been reached in 2024 continuous running, 5 times higher than in \runtwo.
This should allow even more accurate measurements required for constraining various theoretical models~\cite{LHCb:2022ine, LHCb-DP-2022-002}.

Another unique feature of the \lhcb experiment is the ability to carry out studies of quark-gluon plasma simultaneously in two extreme regions of the QCD phase diagram: at high temperature---low density in collider mode, and high density---low temperature in fixed-target mode.
The latter is achieved by injecting a tiny amount of gas into the beam pipe. 
The injection system, called SMOG, was also replaced in 2019--2021, and the new upgraded SMOG2 system~\cite{LHCb-TDR-020} allowed to reach 100 times higher gas density than in \runtwo.

The increased instantaneous luminosity requires special measures for the safety and efficiency of data taking.
To ensure the successful operation of the \lhcb detector, its Beam and Background monitoring systems were also upgraded.
They include the new online luminometer \plume~\cite{CERN-LHCC-2021-002, Barsuk:2743098}, the Beam Condition Monitors (BCM)~\cite{LHCb-DP-2022-002, Ilgner:1233669} dumping the \lhc beams in case of high backgrounds, and the Radiation Monitoring System for \runthree (\rmsthree)~\cite{Chernyshenko:2024}.
The design and operation of the latter is the subject of this paper.

The \rmsthree design is based on the technology of radiation hard Metal Foil Detectors (MFD)~\cite{Pugatch:2004566}.
The signals of the eight MFDs are measured independently of the main \lhcb data acquisition and other subdetectors.
After calibration, they provide a measurement of the luminosity.
A similar value is measured by the main luminometer \plume, and is sent to the \lhc.
During \plume downtime, the \rmsthree luminosity takes over.
The \plume--\rmsthree tandem ensures that the \lhc can steer the beams such that the rate of the interactions per bunch-crossing is kept at the optimal level, which is crucial for the successful detector operation, and prevents excessive irradiation and ageing of the \lhcb components.

\section{The \lhcb experiment}
\label{sec:lhcb}
\subsection{The \lhcb detector}
\label{ssec:lhcb}

\begin{figure}[htbp]
\centering
\includegraphics[width=\textwidth]{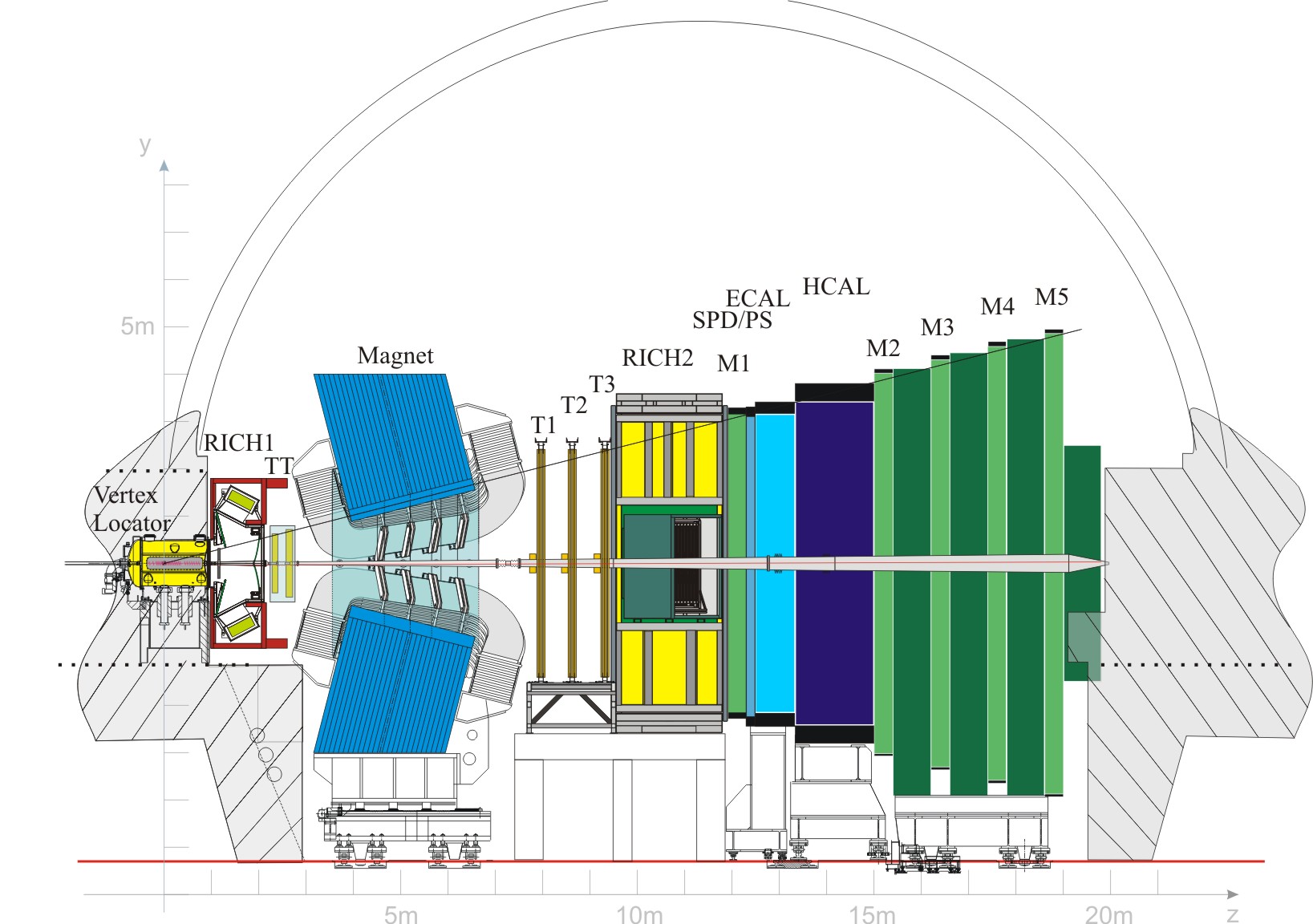}
\caption{
The scheme of the \lhcb detector complex after \upgradeone in 2019--2021 (copied from~\cite{LHCb-DP-2022-002}).
The \lhcb coordinate system is defined with the origin at the interaction point, with the $z$ and $y$ axes pointing to the right and upwards, respectively, while the $x$ axis points into the figure plane, making the system right-handed.
According to the \lhc convention, the beam codirectional (opposite) to the $z$ axis is called beam one (two).
\label{fig:lhcb}
}
\end{figure}

The \lhcb detector~\cite{LHCb-DP-2022-002,LHCb-DP-2008-001,LHCb-DP-2014-002} is a single-arm forward spectrometer covering the pseudo\-rapidity range \mbox{$2\!<\!\eta\!<\!5$} (Figure~\ref{fig:lhcb}). 
The detector has been substantially upgraded prior to the \runthree data-taking period, in order to match the performance of the Run~1--2 detector, while allowing it to operate at approximately five times increased luminosity~\cite{LHCb-DP-2022-002}.
The high-precision tracking system has been fully replaced and consists of a silicon-pixel vertex detector surrounding the interaction region~\cite{LHCb-TDR-013}, a large-area silicon-strip detector~\cite{LHCb-TDR-015} located upstream of a dipole magnet with a bending power of about $4{\mathrm{\,T\,m}}$, and three stations of scintillating-fibre detectors~\cite{LHCb-TDR-015}.
Different types of charged hadrons are distinguished using information from two ring-imaging Cherenkov detectors~\cite{LHCb-DP-2012-003,LHCb-TDR-014}. 
The whole photon detection system of the Cherenkov detectors has been renewed for the upgraded detector. 
Photons, electrons and hadrons are identified by a calorimeter system consisting of electromagnetic and hadronic calorimeters.
Muons are identified by a system composed of alternating layers of iron and multiwire proportional chambers~\cite{LHCb-DP-2012-002}.
Readout of all detectors into an all-software trigger~\cite{LHCb-TDR-016} is a central feature of the upgraded detector, facilitating the reconstruction of events at the maximum \lhc interaction rate, and their selection in real time.
The trigger system is implemented in two stages: a first inclusive stage based primarily on charged particle reconstruction, which reduces the data volume by roughly a factor of 20, and a second stage, which performs the full offline-quality reconstruction and selection of physics signatures.
A large disk buffer is placed between these stages to hold the data while the real-time alignment and calibration are being performed.

\subsection{Beam and background monitoring systems}
\label{ssec:bb}
Stable operation during the \runthree period (2022--2026) in a harsh radiation environment created by hadron collisions in the \lhcb interaction region requires continuous online monitoring of the beam-related conditions, particularly the instantaneous luminosity, the collision patterns of the \lhc beams and the beam-induced background levels.
The system should rapidly react and report any anomalies and abrupt changes of the running conditions.

New technical requirements for the online monitoring systems at \lhcb were redefined for \runthree and resulted in the construction of three independent systems: \plume, BCM and \rmsthree, briefly presented below.
All of them are integrated into the central control system of the \lhcb experiment.

\textbf{\plume} (Probe for LUminosity MEasurement)~\cite{CERN-LHCC-2021-002, Barsuk:2743098} is a new detector system specially developed for measuring the instantaneous luminosity in real time.
In the \plume luminometer, installed upstream wrt the beam one direction of $\sim\!2\m$ of the \lhcb interaction point, 48 photomultipliers count Cherenkov light photons created in quartz radiators.
The detector reports online both the total instantaneous luminosity at \lhcb and the luminosity per bunch-crossing.

\textbf{BCM} (Beam Conditions Monitor)~\cite{LHCb-DP-2022-002, Ilgner:1233669} is a key safety system of the \lhcb experiment, which performs fast control of the radiation levels induced by the \lhc beams. 
It consists of 16 poly-crystalline chemical vapour deposited (pCVD) diamond pad sensors.
They are installed on both sides of the interaction point and read out every $40\mus$.
If the combination of the signals accumulated in any 80 or $1280\mus$ time intervals exceeds the predefined thresholds, the system triggers the \lhc beam dumps.
This guarantees the safe operation and protection of the entire \lhcb detector complex.
The system operated successfully in Run~1--2.
During \upgradeone, the sensors have been replaced, together with the support structures and electronics.

\begin{figure}[htbp]
\centering
\includegraphics[width=0.6\textwidth]{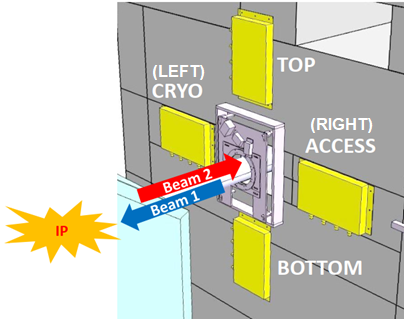}
\caption{
The layout of four \rmsthree detector modules, each with two  detectors.
``Cryo'' and ``Access'' denote the LEFT and RIGHT module pairs when viewed from the \lhcb cavern.
\label{fig:rmsgeo}}
\end{figure}

\textbf{\rmsthree} (Radiation Monitoring System for \runthree)~\cite{Chernyshenko:2024} is a Metal Foil Detectors assembly consisting of four modules mounted 2.2\aunit{m} upstream of the interaction point on the reinforced concrete wall separating the \lhcb cavern from the \lhc tunnel.
They are installed in a ``TOP--BOTTOM'' and ``LEFT--RIGHT'' orientation with respect to the beam pipe, as shown in Figure~\ref{fig:rmsgeo}.
The system measures the \lhcb instantaneous luminosity together with \plume. 
Moreover, the \rmsthree operates autonomously, sending data online to the \lhcb database every second.
This fact belongs to MFD’s operation, which provides an output frequency generated by a charge-to-frequency converter as a response to particle flux.

The \rmsthree with its cross-like layout is dedicated to monitoring the stability of the beam-beam interaction region and background presence, by using the asymmetry method, measuring the asymmetry of responses of its detector pairs (``LEFT--RIGHT'', ``TOP--BOTTOM'').
The asymmetry method involves evaluating each measurement by all \rmsthree detectors using the following equation:  
\[
    A_{\text{\rmsthree}} = \frac{R_{i}-R_{j}}{R_{i}+R_{j}},
\]
here $R_{i}$ and $R_{j}$ are the response frequencies of detectors $i$ and $j$ corresponding to the ``LEFT--RIGHT'' or ``TOP--BOTTOM'' pairs.

Each of the 8 MFD detectors responds to the flux of charged particles originating also from a background.
As far as during the three years of the \runthree operation, the \lhc beams interaction region was stable within $40\mum$ in the plane perpendicular to the beampipe) and taking into account the exclusive \rmsthree stability, we can establish the presence of background by observation of the events localization shifts in 2D histograms of response asymmetries measured by detector pairs (``LEFT--RIGHT'', ``TOP--BOTTOM'') and provide an alarm message to the physicist on duty in the \lhcb Control room.

\section{The \rmsthree Metal Foil Detectors assembly}
\label{sec:rms}
Radiation Monitoring Systems based on the MFD~\cite{Pugatch:2004566} technology are well established.
They have been successfully running in various experiments: in the \lhcb Inner Tracker~\cite{Agari:1026718, Pugatch:2009, Alessio:2015proc, Okhrimenko:1563821, Okhrimenko:2011lli, Pugatch:684677} during Run~1--2 (2009--2018), as a luminosity monitor in the HERA-B experiment~\cite{Pugatch:2002204, Aushev:2001, Riechmann:1996276}, and as almost ``transparent'' beam profile monitors at MPIfK (Tandem generator, Heidelberg)~\cite{Aushev:2001, Riechmann:1996276} and \cern (SPS test line, Geneva, Switzerland)~\cite{Pugatch:684677}.

\begin{figure}[htbp]
\centering
\includegraphics[width=0.79\textwidth]{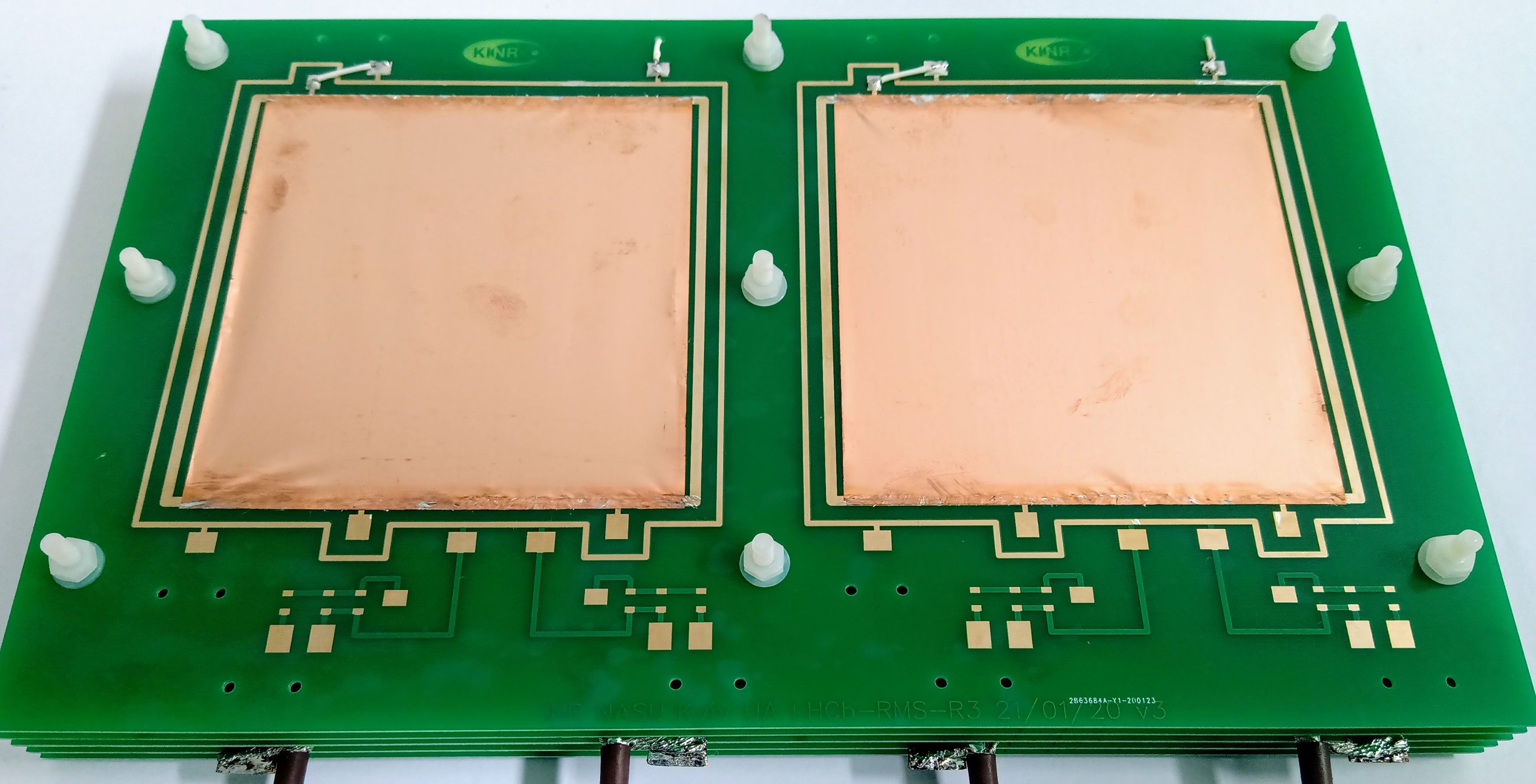}\\
\includegraphics[width=0.79\textwidth]{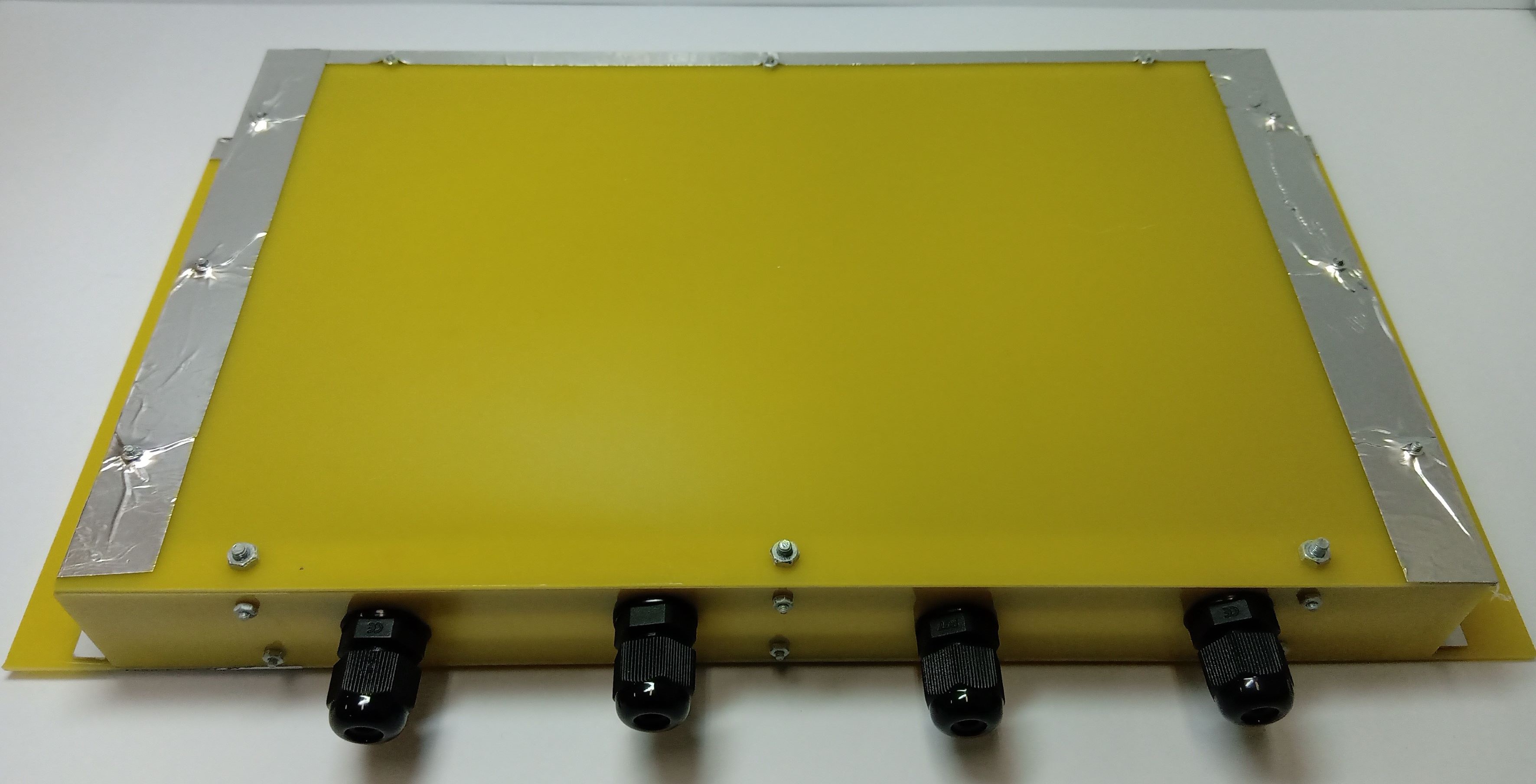}
\caption{
Photos of the detector module.
The top one shows two  detectors mounted on a single PCB with protective rings and contact pads for soldering BNC connectors.
The bottom photo is taken after mounting the detector inside its  G10 case but before covering it with the protective aluminium foil.
Two BNC connectors supply the bias voltage to the collecting foils, and the other two read out the  detector signals, as shown in Figure~\ref{fig:rmsread-a}.
\label{fig:rmsphoto}
}
\end{figure}

\begin{figure}[htbp]
\centering
\includegraphics[width=0.7\textwidth]{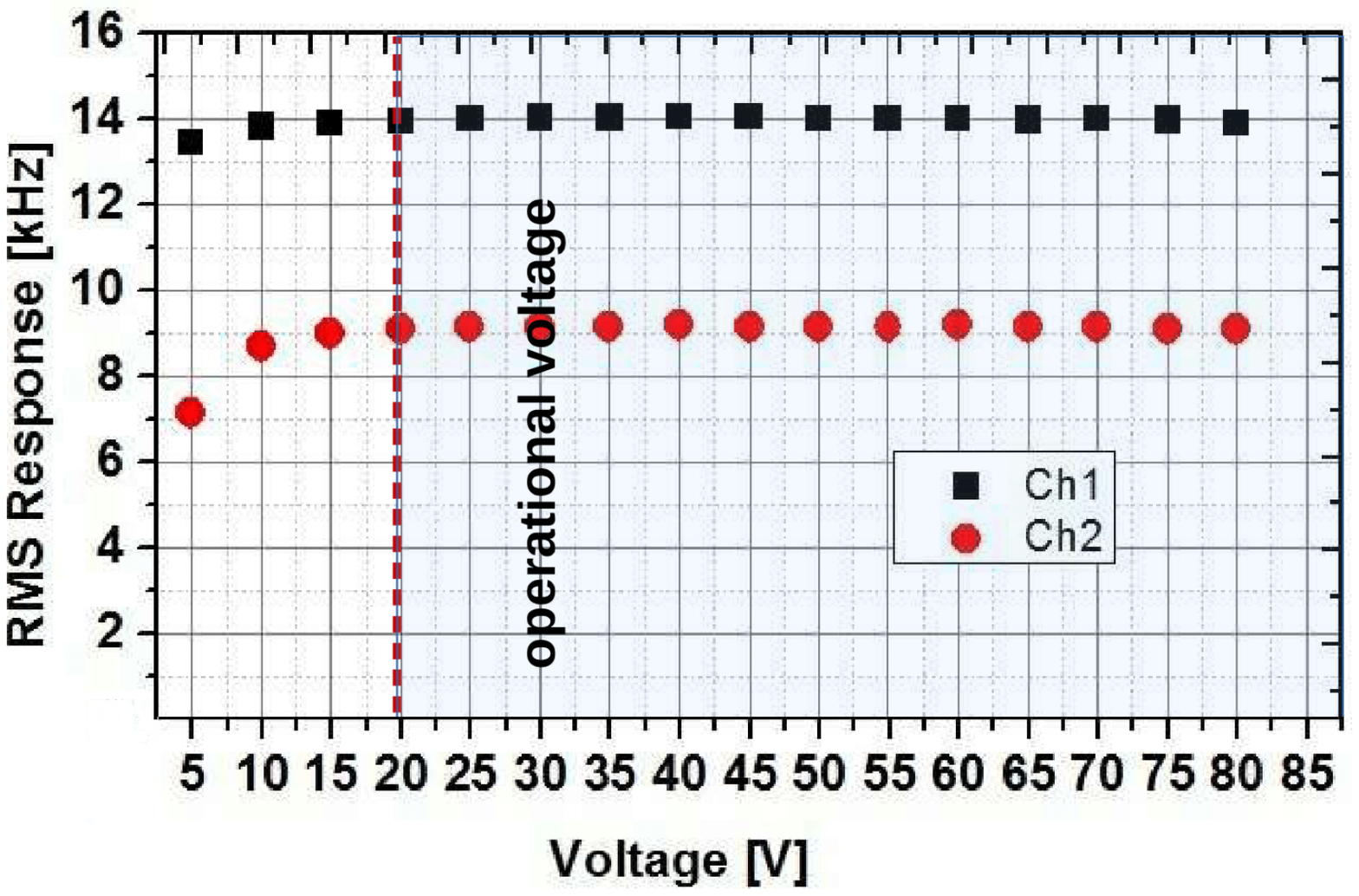}
\vspace*{-1.5em}
\caption{
The response of a pair of MFD  detecrtors irradiated by a $\mathrm{^{90}Sr}$ beta-source on the voltage applied between the sensor and collecting foils.
The \rmsthree  detectors operate at $24\aunit{V}$.
\label{fig:rmsresp1}
}
\end{figure}

The \rmsthree geometry layout (Figure~\ref{fig:rmsgeo}) was designed to provide the monitoring of the \lhcb beam and background.
For reliability, the MFD  detectors are duplicated in each module (Figure~\ref{fig:rmsphoto}).
The final positions of the ones nearest to the beam are chosen for having up to 300\khz signal rate at the nominal \lhcb luminosity of \mbox{$2\!\times\!10^{33}\instlumi$}.
This should provide, on the one hand, $\sim\!0.1\%$ statistical accuracy of the frequency measurement and, on the other hand, a linear response to the instantaneous luminosity.
The non-linearity is expected to appear at $1.3\mhz$ due to detector readout electronics.
The distances from the beam axis to the centres of the two $9\times9\cma$ detectors in each module are 345 and 465\mm.
The installation on the wall was performed with submillimeter accuracy by the technical support group at \lhcb using the metrology methods developed at \cern.

The main goal of this article is to present the \rmsthree capability to monitor the beam and background monitoring in the \lhcb experiment, rather than to measure radiation dose loads. 
While its eight MFDs perform measurement of charged particle fluxes (luminosity, mainly), their specific cross-like array of four pairs (``TOP--BOTTOM'', ``LEFT--RIGHT'') around the beam pipe allows for monitoring the spatial stability of the luminosity origin (interaction region localization) by measuring asymmetries of pairs response.
The high precision of those measurements allows to establish the background contribution if the shifts of the asymmetries exceed the limits determined at background-free conditions.

Detector modules are assembled using polyamide bolts, nuts and washers in a housing made of G10-type material (glass textolite impregnated with an epoxy-phenolic connector), and covered with a thin aluminium foil.
This provides both mechanical protection and electromagnetic shielding.
Figure~\ref{fig:rmsphoto} shows a photo of the detector module without and with protective housing.
All construction materials of the \rmsthree detector modules were chosen following the safety rules in the \lhc experimental halls.

The SEE current increases with the positive voltage bias applied between the sensor and collecting foils, until it reaches a saturation plateau, as shown in Figure~\ref{fig:rmsresp1}.
It was obtained by irradiating a pair of MFD modules with a $\mathrm{^{90}Sr}$ beta-source.

The difference in responses is due to the arbitrary positioning of the source between two sensors to irradiate them simultaneously to save time.

The saturation is reached at $\sim\!15\aunit{V}$, while the \rmsthree is operated 
at 24\aunit{V}.
The response stays constant up to 80\aunit{V}.

Using a $\mathrm{^{90}Sr}$ beta-source at \cern, the calibration factor between the MFD output frequency and the number of particles hitting its area was determined to be \mbox{$550 \pm 180\aunit{particles/detector/\hz}$}, where the $33\%$ error is dominated by the $30\%$ uncertainty of the beta-source activity.
 It should be noted that the primary goal of such measurements was checking the \rmsthree functionality at \cern, so the available beta-source was used.
This and other parameters of the \rmsthree modules are summarized in Table~\ref{tab:rms}.

\begin{table}[htbp]
\centering
\caption{
The main characteristics of the \rmsthree modules.
The MFD are well-known for their radiation tolerance, they easily withstand the harshest radiation levels in the \lhcb environment.
The frequency response of the detector to one incident particle was measured with a $\mathrm{^{90}Sr}$ beta-source as explained in the text.
\label{tab:rms}
}
\smallskip
\begin{tabular}{l|l}
  \hline
  Parameter & Value \\
  \hline
  sensor & $9\!\times\!9$\cma $\times$ 50\mum (0.1\Xrad) Cu foil \\
  operational voltage & 24\aunit{V} \\
  sensor--collecting foil &\\
  capacitance & 25\aunit{pF}  \\
  radiation hardness & $\simeq 1$ \aunit{GGy}, $\simeq 10^{20}$ \mip/\cma \\
  & $10^3\ \times$ expected at \lhcb in \runthree  \\
  calibration &  ($550 \pm 180$) $\mathrm{^{90}Sr}$-$\beta^-$/detector/\hz\\
  \hline
\end{tabular}
\end{table}

\begin{figure}[htbp]
\centering
\subfloat[\label{fig:rmsread-a}]{\includegraphics[width=\textwidth]{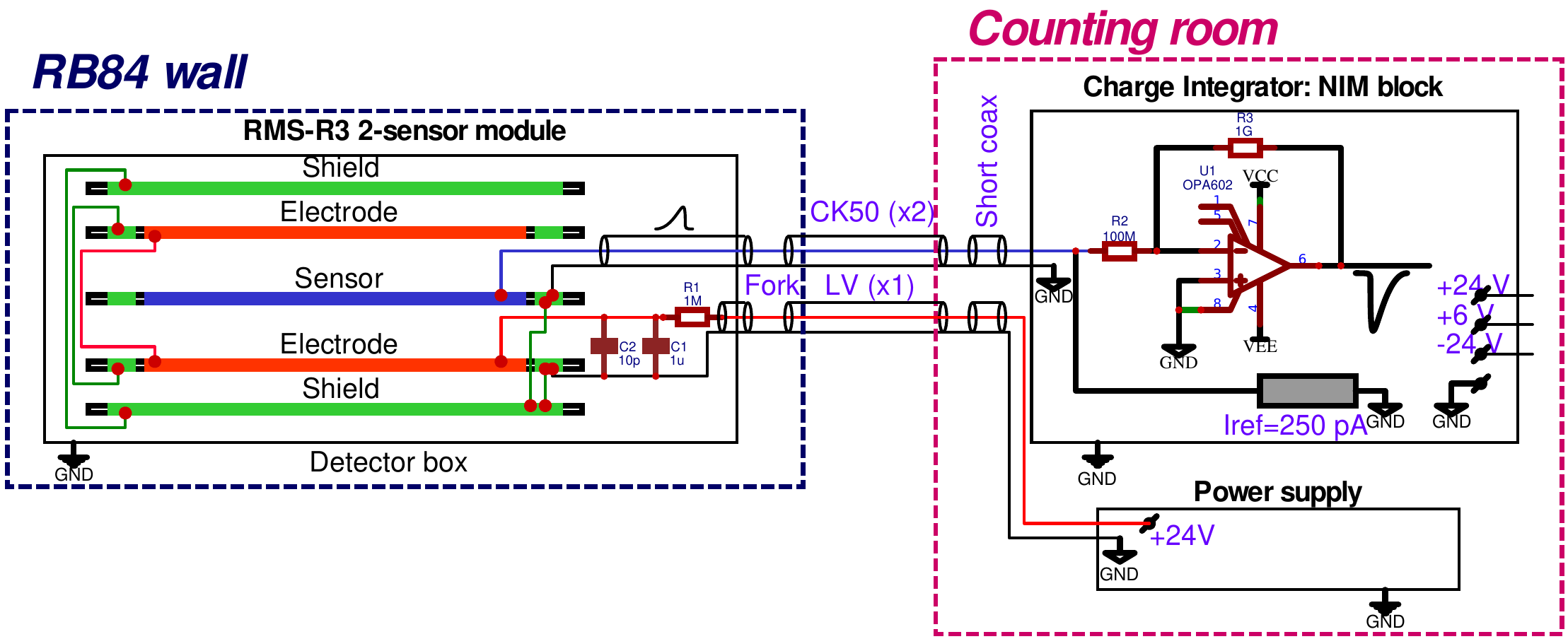}}\\
\subfloat[\label{fig:rmsread-b}]{\includegraphics[width=\textwidth]{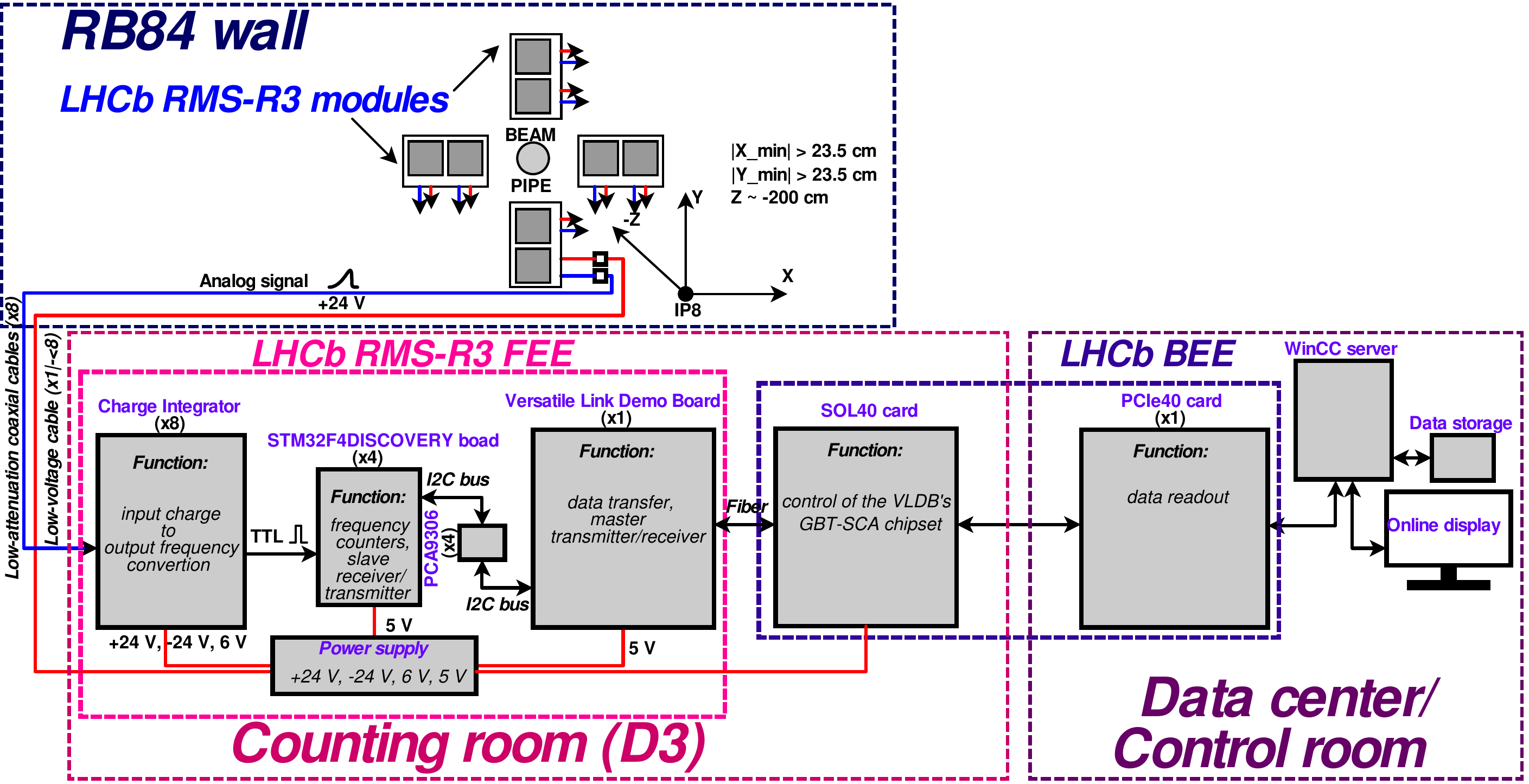}}    
\caption{
(a).
RB84 wall block: the detailed scheme of the \rmsthree detectors mounted on the RB84 wall, illustrating the structure of a single detector module.
Counting room block: the simple schema of Charge Integrator connected by special cables to the \rmsthree detectors located at a $100\m$ distance away in the Counting room.\\
(b).
RB84 wall block: the schematic block diagram of the \rmsthree detector modules.
Counting and Control rooms block: the schematic block diagram of the \rmsthree readout chain, which consists of front-end (FEE) and back-end electronics (BEE), with the devices in the Counting room shown on the left and the Control room shown on the right, respectively.
\label{fig:rmsread}}

\end{figure}

One MFD detector together with the electrical connections is shown schematically in 
Figure~\ref{fig:rmsread-a}. 
The five-layer structure mounted on the printed circuit board has the sensor in the centre, two foils collecting secondary emitted electrons on both sides, and two shielding layers (Figure~\ref{fig:rmsread-a}, RB84 Wall block). 
Charged particles hitting the \mbox{$9\!\times\!9\cma$}, $50\mum$ thick copper foil in the centre generate secondary electron emission (SEE)~\cite{Sternglass:1957, Bruining:2016}.
The induced positive charge is read out by a charge-to-frequency converter VFC110, sensitive at femto-Coulomb level (Figure~\ref{fig:rmsread-a}, Counting room block).
It has a stable voltage source to provide the baseline frequency and has a linear response in the input current range from $10\aunit{fA}$ to $20\aunit{nA}$.
The top and bottom shields of $20\mum$ copper foils ensure stable operation in high electromagnetic and radiation fields.
The impact of environmental conditions, like temperature, humidity, \etc, on the operational characteristics is minimized.
The printed circuit boards are $1\mm$ thick and are made of FR-4 glass fiber with $56\mum$ thick copper tracks and pads.
The contact elements are covered using ENIG-RoHS (\mbox{$\mathrm{Ni}+\mathrm{Au}$}) method.
Special protective rings surround the sensor and collecting foils and isolate them from potential surface cross-talk currents.
The surface-mounted elements of RC filters are soldered on the collecting foil panels.
They filter out high-frequency noise picked up by the power lines.

The chosen module design has the advantage of providing sufficient mechanical rigidity, stable electrical contacts and relatively easy manufacturing and mounting.

The schematic block diagram of the electronics and the readout lines of the \rmsthree detector modules are presented in 
Figure~\ref{fig:rmsread-b}.
As shown in the bottom, the sensor foils in the grounded modules are connected by low attenuation cables to the highly sensitive charge integrators mounted in the NIM crate in the counting room D3, located at 100\m distance.
The charge integrator has a built-in Voltage–Frequency Converter (VFC) and a highly stable current source ($250\aunit{pA}$) providing permanent calibration.
The middle panel represents the front-end and back-end electronics on the left and right sides, respectively.
The front-end electronics consists of the Charge Integrator, STM32-F4DISCOVERY 32-bit counters~\cite{STM:01}, the Versatile Link Demo communication board (VLDB), and the PCIe40 SOL Card for \ecs control~\cite{LHCb-DP-2022-002, Barbosa-Marinho:545306}, all mounted in the \lhcb counting room D3. 
The recorded values are transmitted by the VLDB and PCIe40 boards using 7-bit packets of the $\mathrm{I^2C}$ protocol.
The back-end electronics includes the PCIe40 Card, WinCC system (SCADA) from Siemens~\cite{Cardoso:2015kvo, Barbosa:2295705}, the data storage in the data centre and the displays located in the LHCb Control Room. 

The \rmsthree online monitors present the instantaneous values and the evolution of the luminosity, together with all main parameters of the detectors: their signal frequencies, the bias voltages, and the status of the system components. The data are archived for further reference.

In each charge integrator, its internal 250\aunit{pA} stable current source generates a constant 20--25\khz baseline output.
This is done to continuously check the operating conditions of all eight channels of the \rmsthree system.
After three years of operation, the baseline was stable within $0.5\%$, and its noise did not exceed 20\hz, i.e. $0.02\%$ of the nominal maximal response frequency of 300\khz.

\section{Real-time luminosity monitoring and performance of \rmsthree}
\label{sec:lumi}

The absolute luminosity measurements at \lhcb are described in Refs.~\cite{LHCb:2014, LHCb:2012}.
For any quantity $f$ proportional to the instantaneous luminosity $\lum_{inst}$,
\begin{equation*}
    \lum_{inst} = Af,
\end{equation*}
the proportionality coefficient $A$ can be determined using the van der Meer calibration method~\cite{vanderMeer:296752}.
To illustrate the idea, let's consider the collision of two bunches with $N_{1,2}$ particles and $\rho_{1,2}$ transverse densities moving in opposite directions, and denote by $\lum^{BX}$ {\it the integrated} luminosity {\it per bunch-crossing}.
Then, the corresponding {\it instantaneous} luminosity is 
\begin{equation}
  \lum^{BX}_{inst} = \lum^{BX} f_{LHC} = A f^{BX},
    \label{eq:vdm2}
\end{equation}
where \mbox{$f_{LHC} = 11.245\khz$} is the \lhc bunch revolution frequency and $f^{BX}$ is the rate induced only by the given bunch crossing.
On the other hand, by the definition of the integrated luminosity, 
\begin{equation*} 
\lum^{BX} = N_1 N_2 \iint \rho_1(x,y) \rho_2(x,y) \deriv x\, \deriv y.
\end{equation*}
If the first bunch is separated by
$-\Delta x$, $-\Delta y$ in the transverse plane, the integral over $\Delta x$, $\Delta y$ of this luminosity normalized by the number of particles is
\begin{equation}
\iint \frac{\lum^{BX}(\Delta x,\, \Delta y)}{N_1N_2}\,\deriv\Delta x\,\deriv\Delta y = 
\iint \rho_1(x_2 + \Delta x,\, y_2 + \Delta y) \rho_2(x_2,y_2) \deriv x_2\, \deriv y_2\, \deriv\Delta x\, \deriv\Delta y = 1,
\label{eq:vdm3}
\end{equation}
where $x_2, y_2$ are the coordinates of the stationary second beam. This can be easily seen by introducing new independent variables \mbox{$x_1 = x_2 + \Delta x,\ y_1 = y_2 + \Delta y$} which decouple the integrals \mbox{$\iint \rho_1(x_1,\, y_1)\,\deriv x_1\,\deriv y_1$} and \mbox{$\iint \rho_2(x_2,\, y_2)\,\deriv x_2\,\deriv y_2$}.
The latter are equal to unity due to the normalization of $\rho_{1,2}$.

Using Eqs.~\ref{eq:vdm2} and~\ref{eq:vdm3} one gets the calibration factor
\begin{equation*}
A =
\left( \iint\frac{f^{BX}(\Delta x,\, \Delta y)}{f_{LHC}N_1N_2}\,\deriv\Delta x\,\deriv\Delta y\right)^{-1} =
\left( \iint r^{BX}_{sp}\,\deriv\Delta x\,\deriv\Delta y\right)^{-1}
\end{equation*}
by measuring $f^{BX}(\Delta x,\, \Delta y)$ in $\Delta x$, $\Delta y$ separation scans. Here, 
\begin{equation*}
r^{BX}_{sp} = \frac{f^{BX}}{f_{LHC}N_1N_2}
\end{equation*}
is the normalized or ``specific'' rate ratio.

If the bunch densities are factorizable into $x$ and $y$ independent parts, \mbox{$\rho_{1,2}(x,y) = \rho_{1,2}^x(x) \cdot \rho_{1,2}^y(y)$}, the two-dimensional scan can be reduced to two one-dimensional scans, with only horizontal or vertical beam separations.
In this case,
\begin{equation}
    A = 
    \left( 
    \frac{\int r^{BX}_{sp}(\Delta x, \Delta y_0)\,\deriv\Delta x
        \times \int r^{BX}_{sp}(\Delta x_0, \Delta y)\,\deriv\Delta y}
        {r^{BX}_{sp}(\Delta x_0,\Delta y_0)}
    \right)^{-1}.
\label{eq:vdm6}
\end{equation}

The values $r^{BX}_{sp}(\Delta x, \Delta y_0)$ and $r^{BX}_{sp}(\Delta x_0, \Delta y)$ with constant $\Delta y_0$ and $\Delta x_0$, respectively, are measured in the horizontal and vertical beam separation scans, while $r^{BX}_{sp}(\Delta x_0,\Delta y_0)$ in the denominator is obtained at the intersection of the $x$ and $y$ scan axes.
See Ref.~\cite{Balagura:2021} for the proof of Eq.~\ref{eq:vdm6} and more details on van der Meer method.

In practice, the absolute calibration factor in Eq.~\ref{eq:vdm6} was obtained at \lhcb in van der Meer scans for \plume luminometer, distinguishing the bunches, while \rmsthree signals, integrated over all bunches, were recalibrated with respect to \plume. 

\begin{figure}[htbp]
\centering
\includegraphics[width=0.8\textwidth]{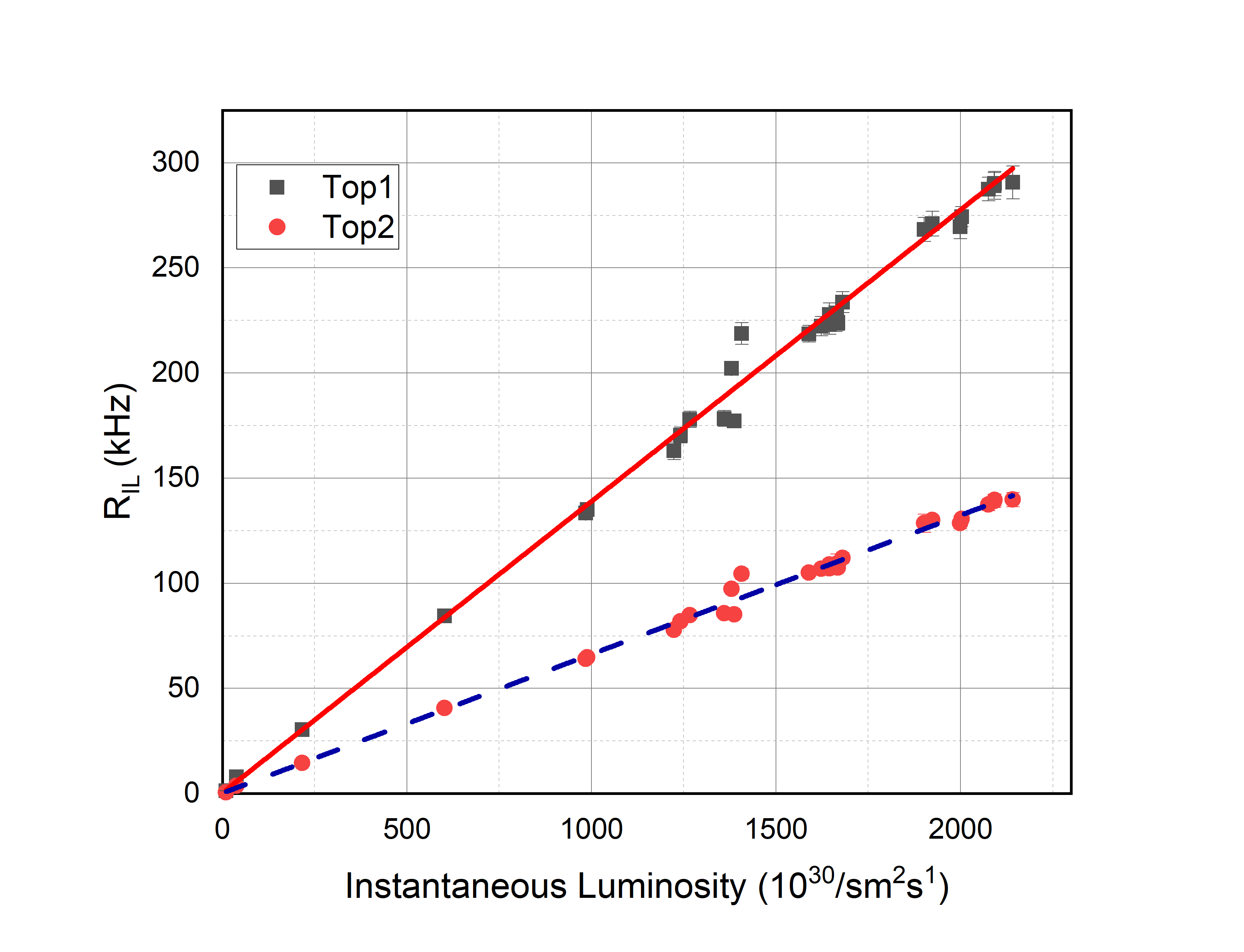}
\vspace*{-1em}
\caption{
The upper (lower) set of points shows the output frequency $R_{IL}$ in $\khz$ of the \rmsthree ``Top'' sensor closer to (farther from) the beam versus 
the instantaneous luminosity measured by \plume, in proton-proton collision data in 2024.
Each point represents the different \lhc fills.  
The solid (dashed) lines represent the linear fits.
\label{fig:rmsresp2}
}
\end{figure}

Figure~\ref{fig:rmsresp2} demonstrates the linearity of the \rmsthree response with respect to the instantaneous luminosity measured by \plume in proton-proton collisions recorded in 2024.
Each point represents the different \lhc fills.
The two lines correspond to two detectors in the ``Top'' module, the one closer to the beam has higher frequencies.
The correlations in other \rmsthree modules are similar.
The evaluated deviations from the linear fits are in the range of $3\%$ for both Top module detectors.
As for other detectors, their responses differ from them by less than $1\khz$ (at $300\khz$). 

The \rmsthree has a precise technique which allows to measure the output frequency of $300\khz$ at nominal instantaneous luminosity of the \lhcb with an accuracy of $20\hz$. That allowed us to observe the $<1 \% $ shifts of the loci of events on a 2D histogram of asymmetries with the evaluation of its centroids with a precision of $0.1 \%$.

The deviations from the linear trend seen in the \rmsthree detectors for luminosities slightly below \mbox{$1500\!\times\!10^{30}$\instlumi} from the average value are a manifestation of background during the fill.
For these fills, it was more recognizable than for the others because during these fills, there was a change in the luminosity level throughout the fill, which was observed by the system.

\begin{figure}[htbp]
\centering
\includegraphics[width=0.7\textwidth]{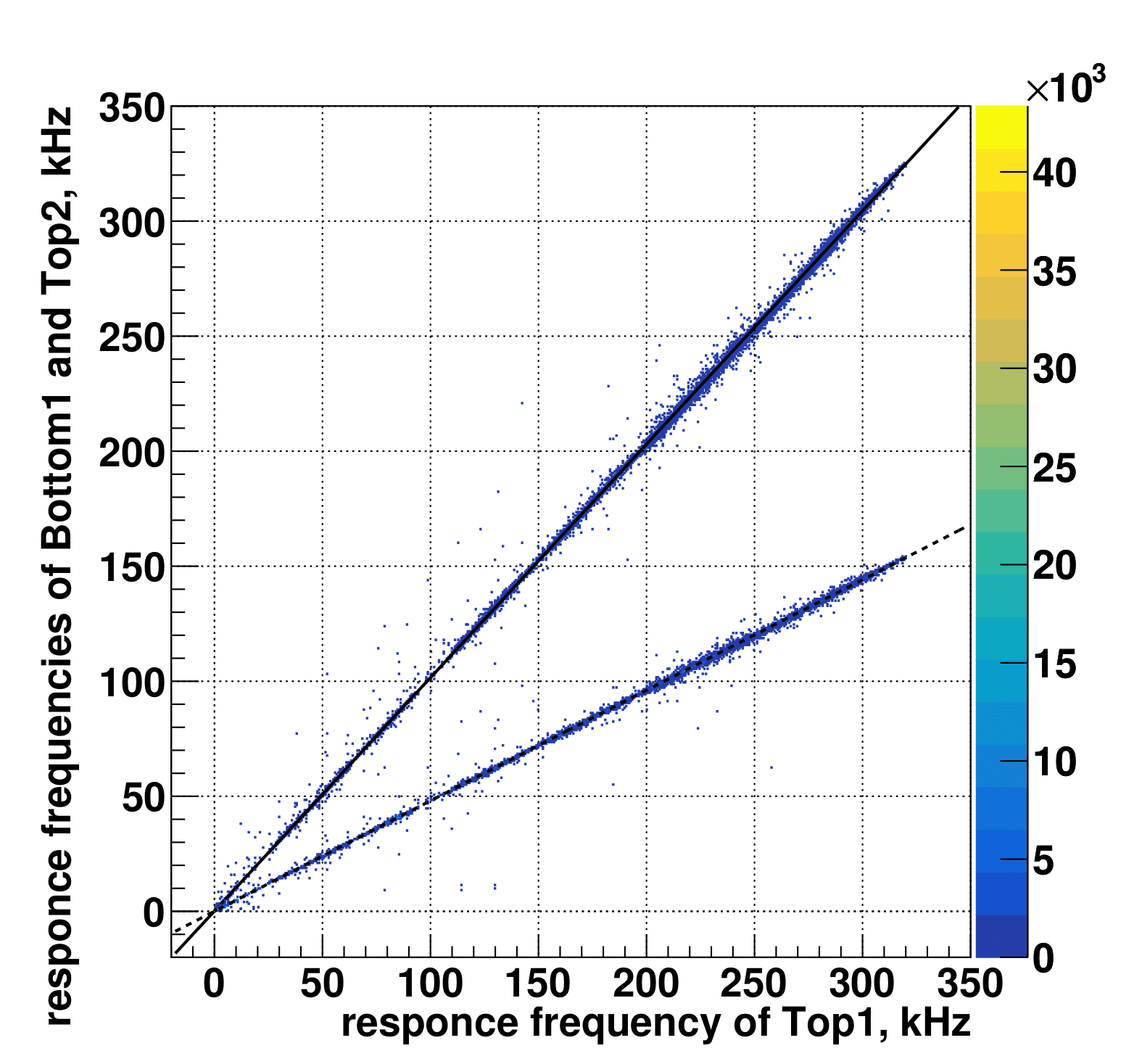}
\caption{
The upper (lower) set of points shows the correlation between the frequencies measured in the closest to the beam detectors in the ``BOTTOM1'' and the ``TOP'' \rmsthree modules (the farthest and closest to the beam-pipe detectors in the ``TOP'' module).
The lines represent the results of the linear fit: 
$\text{slope} = 1.015$ and $\text{offset} = 0.13$
($\text{slope} = 0.481$, $\text{offset} = -0.03$), where the offsets are calculated after a constant baseline subtraction.
Each point represents the \rmsthree response for one second.
All data presented for several fills.
\label{fig:rmsresp3}
}
\end{figure}

The bottom line in Figure~\ref{fig:rmsresp3} shows the linearity of the ``TOP'' detectors responses with respect to each other.
The upper line correlates the closest to the beam detectors in the ``TOP'' and ``BOTTOM'' modules. For other pairs of detectors, the correlations are also similar. 
It is worthwhile to notice that the number of events for the outliers is 4--6 orders of magnitude lower than for points in the center of the fitting line.

This demonstrates the \rmsthree capability to deliver high-precision real-time luminosity monitoring over a wide dynamic range.
The reproducibility of the measurements across multiple sensing points confirms the robustness of the design and the reliability of the system for long-term operation. 
The high degree of stability observed over extended periods allows \rmsthree to provide continuous and accurate information on radiation conditions in the \lhcb interaction region, which is crucial for the beam and background monitoring.

\begin{figure}[htbp]
\centering
\includegraphics[width=\textwidth]{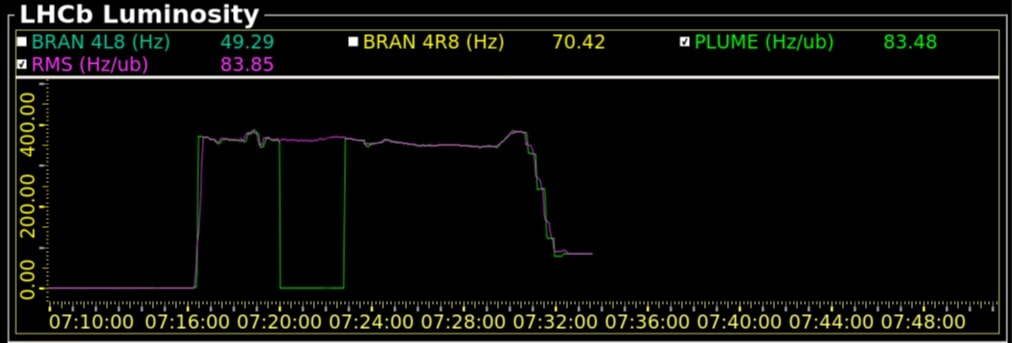}
\caption{
An example of the time evolution of the instantaneous luminosity measured by \rmsthree and the main online luminometer \plume, shown by the purple and green lines, respectively. In the period of \plume downtime, when its values drop to zero, \rmsthree takes over and becomes the main luminometer of LHCb.
\label{fig:rmsdisp}
}
\end{figure}

 Both \plume and \rmsthree signals are provided to the \lhcb Control Room and are available in a single window where the luminosity is shown to the \lhcb operators. The time evolution of the luminosities measured by \rmsthree and the main online luminometer \plume during one fill as an example is shown in Figure~\ref{fig:rmsdisp}. Such plots are always available in the \lhcb control room online.  It is important to emphasize, the \plume provides the absolute luminosity value, and the \rmsthree is calibrated by it.  

Only one instantaneous luminosity signal is provided to the \lhc machine, and it is chosen based on a hierarchy where \plume is first and \rmsthree is the first back-up system. Therefore, RMS is effectively the first redundant luminometer in \lhcb. The \lhc uses in first place the signal from the experiment and only as a back-up solution, the BRANs~\cite{Matis:2016raz}.

The calibrated luminosity provided by the \plume--\rmsthree tandem is used by the \lhc to keep the \lhcb luminosity at the optimal level.
When the \lhc declares ``stable beams'' and the detectors are allowed to take data, the beam intensities in the beginning are high.
Therefore, the \lhc slightly separates the beams at \lhcb.
Then, due to the continuous loss of particles, mainly because of the interactions in the four \lhc experiments, the beams get closer, and at the end of long runs collide head-on.
This procedure is called the luminosity-leveling and is fully automated.
It ensures the maximal event rate which can be accepted by the \lhcb experiment.
The complementarity and redundancy of the \plume and \rmsthree luminosity measurements ensure the accuracy and $100\%$ availability of the luminosity values in real-time, which guarantees that the \lhcb detector always works in the optimal conditions and takes good quality data.

\section{Conclusions and outlook}
\label{sec:con}
The \rmsthree detector assembly provides an autonomous real-time measurement of the instantaneous luminosity in the \lhcb experiment during \runthree.
\rmsthree data demonstrate its linear response with high reproducibility in a five-decade dynamic range of instantaneous luminosity over a long period of operation.
The Metal Foil Detector technology of the \rmsthree is promising for any radiation-hard application like the monitoring of the beam, background and safety conditions in high energy physics experiments.
The MFD detectors can withstand extreme levels of radiation expected at HL-LHC and  future colliders like FCC, and are well suited for the luminosity measurements.

\section*{Acknowledgments}

We'd like to thank Gloria, Matthias, Eric, Pascal, Anna, Augusto, Gennady, Dmytro and Jacky for their invaluable help in implementing the \rmsthree at the \lhcb experiment.

This project has received funding through the EURIZON project, which is funded by the European Union under grant agreement No.871072. Grant \#3014.

\addcontentsline{toc}{section}{References}
\bibliographystyle{LHCb}
\bibliography{main}

\ifx\mcitethebibliography\mciteundefinedmacro
\PackageError{LHCb.bst}{mciteplus.sty has not been loaded}
{This bibstyle requires the use of the mciteplus package.}\fi
\providecommand{\href}[2]{#2}
\begin{mcitethebibliography}{10}
\mciteSetBstSublistMode{n}
\mciteSetBstMaxWidthForm{subitem}{\alph{mcitesubitemcount})}
\mciteSetBstSublistLabelBeginEnd{\mcitemaxwidthsubitemform\space}
{\relax}{\relax}

\bibitem{LHCb-PAPER-2012-031}
LHCb collaboration, {R.\ Aaij \emph{et al.\ }, and A.\ Bharucha} {\em et~al.}, \ifthenelse{\boolean{articletitles}}{\emph{{Implications of LHCb measurements and future prospects}}, }{}\href{https://doi.org/10.1140/epjc/s10052-013-2373-2}{Eur.\ Phys.\ J.\  \textbf{C73} (2013) 2373}, \href{http://arxiv.org/abs/1208.3355}{{\normalfont\ttfamily arXiv:1208.3355}}\relax
\mciteBstWouldAddEndPuncttrue
\mciteSetBstMidEndSepPunct{\mcitedefaultmidpunct}
{\mcitedefaultendpunct}{\mcitedefaultseppunct}\relax
\EndOfBibitem
\bibitem{LHCb:2022ine}
LHCb collaboration, R.~Aaij {\em et~al.},  \ifthenelse{\boolean{articletitles}}{\emph{{Future physics potential of LHCb}}}{}, \href{https://cds.cern.ch/record/2806113}{LHCb-PUB-2022-012, CERN-LHCb-PUB-2022-012}, CERN, Geneva, 2022\relax
\mciteBstWouldAddEndPuncttrue
\mciteSetBstMidEndSepPunct{\mcitedefaultmidpunct}
{\mcitedefaultendpunct}{\mcitedefaultseppunct}\relax
\EndOfBibitem
\bibitem{LHCb-DP-2022-002}
LHCb collaboration, R.~Aaij {\em et~al.}, \ifthenelse{\boolean{articletitles}}{\emph{{The LHCb Upgrade I}}, }{}\href{https://doi.org/10.1088/1748-0221/19/05/P05065}{{JINST} \textbf{19} (2024) P05065}, \href{http://arxiv.org/abs/2305.10515}{{\normalfont\ttfamily arXiv:2305.10515}}\relax
\mciteBstWouldAddEndPuncttrue
\mciteSetBstMidEndSepPunct{\mcitedefaultmidpunct}
{\mcitedefaultendpunct}{\mcitedefaultseppunct}\relax
\EndOfBibitem
\bibitem{LHCb-TDR-020}
LHCb collaboration, R.~Aaij {\em et~al.},  \ifthenelse{\boolean{articletitles}}{\emph{{LHCb SMOG Upgrade}}}{}, \href{https://cds.cern.ch/record/2673690}{CERN-LHCC-2019-005}, CERN, Geneva, 2019.
\newblock doi:~\href{https://doi.org/10.17181/CERN.SAQC.EOWH}{10.17181/CERN.SAQC.EOWH}\relax
\mciteBstWouldAddEndPuncttrue
\mciteSetBstMidEndSepPunct{\mcitedefaultmidpunct}
{\mcitedefaultendpunct}{\mcitedefaultseppunct}\relax
\EndOfBibitem
\bibitem{CERN-LHCC-2021-002}
LHCb collaboration, S.~Barsuk {\em et~al.},  \ifthenelse{\boolean{articletitles}}{\emph{{LHCb PLUME: Probe for LUminosity MEasurement}}}{}, \href{https://cds.cern.ch/record/2750034}{CERN-LHCC-2021-002, LHCB-TDR-022}, CERN, Geneva, 2021.
\newblock doi:~\href{https://doi.org/10.17181/CERN.WLU0.M37F}{10.17181/CERN.WLU0.M37F}\relax
\mciteBstWouldAddEndPuncttrue
\mciteSetBstMidEndSepPunct{\mcitedefaultmidpunct}
{\mcitedefaultendpunct}{\mcitedefaultseppunct}\relax
\EndOfBibitem
\bibitem{Barsuk:2743098}
S.~Barsuk {\em et~al.},  \ifthenelse{\boolean{articletitles}}{\emph{{Probe for LUminosity MEasurement in LHCb}}}{}, \href{https://cds.cern.ch/record/2743098}{LHCb-PUB-2020-008, CERN-LHCb-PUB-2020-008}, CERN, Geneva, 2020\relax
\mciteBstWouldAddEndPuncttrue
\mciteSetBstMidEndSepPunct{\mcitedefaultmidpunct}
{\mcitedefaultendpunct}{\mcitedefaultseppunct}\relax
\EndOfBibitem
\bibitem{Ilgner:1233669}
C.~Ilgner {\em et~al.}, \ifthenelse{\boolean{articletitles}}{\emph{{The Beam Conditions Monitor of the LHCb Experiment}}}{} , CERN, Geneva, 2010.
\newblock Comments: Index Terms: Accelerator measurement systems, CVD, Diamond, Radiation detectors\relax
\mciteBstWouldAddEndPuncttrue
\mciteSetBstMidEndSepPunct{\mcitedefaultmidpunct}
{\mcitedefaultendpunct}{\mcitedefaultseppunct}\relax
\EndOfBibitem
\bibitem{Chernyshenko:2024}
S.~B. Chernyshenko, V.~M. Dobishuk, and V.~M. Pugatch, \ifthenelse{\boolean{articletitles}}{\emph{{Functionality features of the RMS-R3 system in the third physics run of the LHCb experiment}}, }{}\href{https://doi.org/10.15407/jnpae2024.02.188}{Nuclear Physics and Atomic Energy \textbf{25} (2024) 188}\relax
\mciteBstWouldAddEndPuncttrue
\mciteSetBstMidEndSepPunct{\mcitedefaultmidpunct}
{\mcitedefaultendpunct}{\mcitedefaultseppunct}\relax
\EndOfBibitem
\bibitem{Pugatch:2004566}
V.~Pugatch {\em et~al.}, \ifthenelse{\boolean{articletitles}}{\emph{Metal foil detectors and their applications}, }{}\href{https://doi.org/doi.org/10.1016/j.nima.2004.07.279}{Nucl.\ Instrum.\ and Methods in Phys.\ Res.\ A \textbf{535} (2004) 566}, Proceedings of the 10th International Vienna Conference on Instrumentation\relax
\mciteBstWouldAddEndPuncttrue
\mciteSetBstMidEndSepPunct{\mcitedefaultmidpunct}
{\mcitedefaultendpunct}{\mcitedefaultseppunct}\relax
\EndOfBibitem
\bibitem{LHCb-DP-2008-001}
LHCb collaboration, A.~A. Alves~Jr.\ {\em et~al.}, \ifthenelse{\boolean{articletitles}}{\emph{{The \lhcb detector at the LHC}}, }{}\href{https://doi.org/10.1088/1748-0221/3/08/S08005}{JINST \textbf{3} (2008) S08005}\relax
\mciteBstWouldAddEndPuncttrue
\mciteSetBstMidEndSepPunct{\mcitedefaultmidpunct}
{\mcitedefaultendpunct}{\mcitedefaultseppunct}\relax
\EndOfBibitem
\bibitem{LHCb-DP-2014-002}
LHCb collaboration, R.~Aaij {\em et~al.}, \ifthenelse{\boolean{articletitles}}{\emph{{LHCb detector performance}}, }{}\href{https://doi.org/10.1142/S0217751X15300227}{Int.\ J.\ Mod.\ Phys.\  \textbf{A30} (2015) 1530022}, \href{http://arxiv.org/abs/1412.6352}{{\normalfont\ttfamily arXiv:1412.6352}}\relax
\mciteBstWouldAddEndPuncttrue
\mciteSetBstMidEndSepPunct{\mcitedefaultmidpunct}
{\mcitedefaultendpunct}{\mcitedefaultseppunct}\relax
\EndOfBibitem
\bibitem{LHCb-TDR-013}
LHCb collaboration, R.~Aaij {\em et~al.}, \ifthenelse{\boolean{articletitles}}{\emph{{LHCb VELO Upgrade Technical Design Report}}}{}  {CERN-LHCC-2013-021}, CERN, Geneva, 2013\relax
\mciteBstWouldAddEndPuncttrue
\mciteSetBstMidEndSepPunct{\mcitedefaultmidpunct}
{\mcitedefaultendpunct}{\mcitedefaultseppunct}\relax
\EndOfBibitem
\bibitem{LHCb-TDR-015}
LHCb collaboration, R.~Aaij {\em et~al.}, \ifthenelse{\boolean{articletitles}}{\emph{{LHCb Tracker Upgrade Technical Design Report}}}{}  {CERN-LHCC-2014-001}, CERN, Geneva, 2014\relax
\mciteBstWouldAddEndPuncttrue
\mciteSetBstMidEndSepPunct{\mcitedefaultmidpunct}
{\mcitedefaultendpunct}{\mcitedefaultseppunct}\relax
\EndOfBibitem
\bibitem{LHCb-DP-2012-003}
M.~Adinolfi {\em et~al.}, \ifthenelse{\boolean{articletitles}}{\emph{{Performance of the \lhcb RICH detector at the LHC}}, }{}\href{https://doi.org/10.1140/epjc/s10052-013-2431-9}{Eur.\ Phys.\ J.\  \textbf{C73} (2013) 2431}, \href{http://arxiv.org/abs/1211.6759}{{\normalfont\ttfamily arXiv:1211.6759}}\relax
\mciteBstWouldAddEndPuncttrue
\mciteSetBstMidEndSepPunct{\mcitedefaultmidpunct}
{\mcitedefaultendpunct}{\mcitedefaultseppunct}\relax
\EndOfBibitem
\bibitem{LHCb-TDR-014}
LHCb collaboration, R.~Aaij {\em et~al.}, \ifthenelse{\boolean{articletitles}}{\emph{{LHCb PID Upgrade Technical Design Report}}}{}  {CERN-LHCC-2013-022}, CERN, Geneva, 2013\relax
\mciteBstWouldAddEndPuncttrue
\mciteSetBstMidEndSepPunct{\mcitedefaultmidpunct}
{\mcitedefaultendpunct}{\mcitedefaultseppunct}\relax
\EndOfBibitem
\bibitem{LHCb-DP-2012-002}
A.~A. Alves~Jr.\ {\em et~al.}, \ifthenelse{\boolean{articletitles}}{\emph{{Performance of the LHCb muon system}}, }{}\href{https://doi.org/10.1088/1748-0221/8/02/P02022}{JINST \textbf{8} (2013) P02022}, \href{http://arxiv.org/abs/1211.1346}{{\normalfont\ttfamily arXiv:1211.1346}}\relax
\mciteBstWouldAddEndPuncttrue
\mciteSetBstMidEndSepPunct{\mcitedefaultmidpunct}
{\mcitedefaultendpunct}{\mcitedefaultseppunct}\relax
\EndOfBibitem
\bibitem{LHCb-TDR-016}
LHCb collaboration, R.~Aaij {\em et~al.}, \ifthenelse{\boolean{articletitles}}{\emph{{LHCb Trigger and Online Upgrade Technical Design Report}}}{}  {CERN-LHCC-2014-016}, CERN, Geneva, 2014\relax
\mciteBstWouldAddEndPuncttrue
\mciteSetBstMidEndSepPunct{\mcitedefaultmidpunct}
{\mcitedefaultendpunct}{\mcitedefaultseppunct}\relax
\EndOfBibitem
\bibitem{Agari:1026718}
M.~Agari {\em et~al.},  \ifthenelse{\boolean{articletitles}}{\emph{{Radiation Monitoring System for the LHCb Inner Tracker}}}{}, \href{https://cds.cern.ch/record/1026718}{LHCb-2007-062, CERN-LHCb-2007-062}, CERN, Geneva, 2007\relax
\mciteBstWouldAddEndPuncttrue
\mciteSetBstMidEndSepPunct{\mcitedefaultmidpunct}
{\mcitedefaultendpunct}{\mcitedefaultseppunct}\relax
\EndOfBibitem
\bibitem{Pugatch:2009}
V.~M. Pugatch {\em et~al.}, \ifthenelse{\boolean{articletitles}}{\emph{{Radiation Monitoring System for LHCb Inner Tracker}}, }{}Ukrainian Journal of Physics \textbf{54} (2009) 418\relax
\mciteBstWouldAddEndPuncttrue
\mciteSetBstMidEndSepPunct{\mcitedefaultmidpunct}
{\mcitedefaultendpunct}{\mcitedefaultseppunct}\relax
\EndOfBibitem
\bibitem{Alessio:2015proc}
O.~Okhrimenko, S.~Barsuk, F.~Alessio, and V.~Pugatch, \ifthenelse{\boolean{articletitles}}{\emph{{LHCb RMS status and operation at 13 TeV}}, }{} in {\em {Proceedings of the third French-Ukrainian workshop on the instrumentation developments for HEP}}, 61--65, 2015, \href{http://arxiv.org/abs/1512.07393}{{\normalfont\ttfamily arXiv:1512.07393}}\relax
\mciteBstWouldAddEndPuncttrue
\mciteSetBstMidEndSepPunct{\mcitedefaultmidpunct}
{\mcitedefaultendpunct}{\mcitedefaultseppunct}\relax
\EndOfBibitem
\bibitem{Okhrimenko:1563821}
O.~Okhrimenko {\em et~al.}, \ifthenelse{\boolean{articletitles}}{\emph{{The Radiation Monitoring System for the LHCb Inner Tracker}}, }{}Conf.\ Proc.\  \textbf{C111010} (2011) WEPMU024\relax
\mciteBstWouldAddEndPuncttrue
\mciteSetBstMidEndSepPunct{\mcitedefaultmidpunct}
{\mcitedefaultendpunct}{\mcitedefaultseppunct}\relax
\EndOfBibitem
\bibitem{Okhrimenko:2011lli}
O.~Y. Okhrimenko, V.~M. Iakovenko, and V.~M. Pugatch, \ifthenelse{\boolean{articletitles}}{\emph{{The first LHC beam impact measured by the LHCb inner tracker radiation monitoring system}}, }{} in {\em {3rd International Conference on Current Problems in Nuclear Physics and Atomic Energy}}, 639--643, 2011\relax
\mciteBstWouldAddEndPuncttrue
\mciteSetBstMidEndSepPunct{\mcitedefaultmidpunct}
{\mcitedefaultendpunct}{\mcitedefaultseppunct}\relax
\EndOfBibitem
\bibitem{Pugatch:684677}
V.~Pugatch {\em et~al.},  \ifthenelse{\boolean{articletitles}}{\emph{{Radiation monitoring system for the LHCb Inner Tracker}}}{}, \href{https://cds.cern.ch/record/684677}{LHCb-2002-067, CERN-LHCb-2002-067}, CERN, Geneva, 2002\relax
\mciteBstWouldAddEndPuncttrue
\mciteSetBstMidEndSepPunct{\mcitedefaultmidpunct}
{\mcitedefaultendpunct}{\mcitedefaultseppunct}\relax
\EndOfBibitem
\bibitem{Pugatch:2002204}
V.~Pugatch, K.~T. Knöpfle, and Y.~Vassiliyev, \ifthenelse{\boolean{articletitles}}{\emph{{Beam profile imaging target}}, }{}\href{https://doi.org/https://doi.org/10.1016/S0375-9474(01)01575-5}{Nuclear Physics A \textbf{701} (2002) 204}, 5th International Conference on Radioactive Nuclear Beams\relax
\mciteBstWouldAddEndPuncttrue
\mciteSetBstMidEndSepPunct{\mcitedefaultmidpunct}
{\mcitedefaultendpunct}{\mcitedefaultseppunct}\relax
\EndOfBibitem
\bibitem{Aushev:2001}
V.~Aushev {\em et~al.}, \ifthenelse{\boolean{articletitles}}{\emph{{Target monitoring system for HERA-B experiment. Multitarget operation}}, }{}Nuclear Physics and Atomic Energy \textbf{2} (2001) 56\relax
\mciteBstWouldAddEndPuncttrue
\mciteSetBstMidEndSepPunct{\mcitedefaultmidpunct}
{\mcitedefaultendpunct}{\mcitedefaultseppunct}\relax
\EndOfBibitem
\bibitem{Riechmann:1996276}
K.~Riechmann, K.~T. Knöpfle, and V.~M. Pugatch, \ifthenelse{\boolean{articletitles}}{\emph{{Pion and proton induced radiation damage to silicon detectors}}, }{}\href{https://doi.org/https://doi.org/10.1016/0168-9002(95)01408-X}{Nucl.\ Instrum.\ and Methods in Phys.\ Res.\ A \textbf{377} (1996) 276}, Proceedings of the Seventh European Symposium on Semiconductor\relax
\mciteBstWouldAddEndPuncttrue
\mciteSetBstMidEndSepPunct{\mcitedefaultmidpunct}
{\mcitedefaultendpunct}{\mcitedefaultseppunct}\relax
\EndOfBibitem
\bibitem{Sternglass:1957}
E.~J. Sternglass, \ifthenelse{\boolean{articletitles}}{\emph{{Theory of Secondary Electron Emission by High-Speed Ions}}, }{}\href{https://doi.org/10.1103/PhysRev.108.1}{Phys.\ Rev.\  \textbf{108} (1957) 1}\relax
\mciteBstWouldAddEndPuncttrue
\mciteSetBstMidEndSepPunct{\mcitedefaultmidpunct}
{\mcitedefaultendpunct}{\mcitedefaultseppunct}\relax
\EndOfBibitem
\bibitem{Bruining:2016}
{H.\ Bruining}, {\em {Physics and Applications of Secondary Electron Emission: Electronics and Waves---a Series of Monographs}}, Elsevier, 2016\relax
\mciteBstWouldAddEndPuncttrue
\mciteSetBstMidEndSepPunct{\mcitedefaultmidpunct}
{\mcitedefaultendpunct}{\mcitedefaultseppunct}\relax
\EndOfBibitem
\bibitem{STM:01}
{STMicroelectronics}, {\em {Reference manual RM0090. STM32F405/415, STM32F407/417STM32F427/437 and STM32F429/439 advanced Arm-based 32-bit MCUs}}, 2024\relax
\mciteBstWouldAddEndPuncttrue
\mciteSetBstMidEndSepPunct{\mcitedefaultmidpunct}
{\mcitedefaultendpunct}{\mcitedefaultseppunct}\relax
\EndOfBibitem
\bibitem{Barbosa-Marinho:545306}
LHCb collaboration, P.~R. Barbosa-Marinho {\em et~al.}, {\em {LHCb online system, data acquisition and experiment control}}, Technical design report. LHCb, CERN, Geneva, 2001\relax
\mciteBstWouldAddEndPuncttrue
\mciteSetBstMidEndSepPunct{\mcitedefaultmidpunct}
{\mcitedefaultendpunct}{\mcitedefaultseppunct}\relax
\EndOfBibitem
\bibitem{Cardoso:2015kvo}
L.~Cardoso {\em et~al.}, \ifthenelse{\boolean{articletitles}}{\emph{{A Framework for Hardware Integration in the LHCb Experiment Control System}}, }{} in {\em {15th International Conference on Accelerator and Large Experimental Physics Control Systems}}, \href{https://doi.org/10.18429/JACoW-ICALEPCS2015-MOPGF052}{ MOPGF052, 2015}\relax
\mciteBstWouldAddEndPuncttrue
\mciteSetBstMidEndSepPunct{\mcitedefaultmidpunct}
{\mcitedefaultendpunct}{\mcitedefaultseppunct}\relax
\EndOfBibitem
\bibitem{Barbosa:2295705}
J.~Barbosa {\em et~al.}, \ifthenelse{\boolean{articletitles}}{\emph{{A Monitoring System for the LHCb Data Flow}}, }{}\href{https://doi.org/10.1109/TNS.2017.2660398}{IEEE Trans.\ Nucl.\ Sci.\  \textbf{64} (2017) 1191}\relax
\mciteBstWouldAddEndPuncttrue
\mciteSetBstMidEndSepPunct{\mcitedefaultmidpunct}
{\mcitedefaultendpunct}{\mcitedefaultseppunct}\relax
\EndOfBibitem
\bibitem{LHCb:2014}
LHCb collaboration, R.~Aaij {\em et~al.}, \ifthenelse{\boolean{articletitles}}{\emph{{Precision luminosity measurements at LHCb}}, }{}\href{https://doi.org/10.1088/1748-0221/9/12/P12005}{Journal of Instrumentation \textbf{9} (2014) P12005}, \href{http://arxiv.org/abs/1410.0149}{{\normalfont\ttfamily arXiv:1410.0149}}\relax
\mciteBstWouldAddEndPuncttrue
\mciteSetBstMidEndSepPunct{\mcitedefaultmidpunct}
{\mcitedefaultendpunct}{\mcitedefaultseppunct}\relax
\EndOfBibitem
\bibitem{LHCb:2012}
LHCb collaboration, R.~Aaij {\em et~al.}, \ifthenelse{\boolean{articletitles}}{\emph{{Absolute luminosity measurements with the LHCb detector at the LHC}}, }{}\href{https://doi.org/10.1088/1748-0221/7/01/P01010}{Journal of Instrumentation \textbf{7} (2012) P01010}, \href{http://arxiv.org/abs/1110.2866}{{\normalfont\ttfamily arXiv:1110.2866}}\relax
\mciteBstWouldAddEndPuncttrue
\mciteSetBstMidEndSepPunct{\mcitedefaultmidpunct}
{\mcitedefaultendpunct}{\mcitedefaultseppunct}\relax
\EndOfBibitem
\bibitem{vanderMeer:296752}
S.~van~der Meer, \ifthenelse{\boolean{articletitles}}{\emph{{Calibration of the effective beam height in the ISR}}}{} , CERN, Geneva, 1968\relax
\mciteBstWouldAddEndPuncttrue
\mciteSetBstMidEndSepPunct{\mcitedefaultmidpunct}
{\mcitedefaultendpunct}{\mcitedefaultseppunct}\relax
\EndOfBibitem
\bibitem{Balagura:2021}
V.~Balagura, \ifthenelse{\boolean{articletitles}}{\emph{{Van der Meer scan luminosity measurement and beam–beam correction}}, }{}\href{https://doi.org/10.1140/epjc/s10052-021-08837-y}{Eur.\ Phys.\ J.\ C \textbf{81} (2021) 26}, \href{http://arxiv.org/abs/2012.07752}{{\normalfont\ttfamily arXiv:2012.07752}}\relax
\mciteBstWouldAddEndPuncttrue
\mciteSetBstMidEndSepPunct{\mcitedefaultmidpunct}
{\mcitedefaultendpunct}{\mcitedefaultseppunct}\relax
\EndOfBibitem
\bibitem{Matis:2016raz}
H.~S. Matis {\em et~al.}, \ifthenelse{\boolean{articletitles}}{\emph{{The BRAN luminosity detectors for the LHC}}, }{}\href{https://doi.org/10.1016/j.nima.2016.12.019}{Nucl.\ Instrum.\ Meth.\ A \textbf{848} (2017) 114}, \href{http://arxiv.org/abs/1612.01238}{{\normalfont\ttfamily arXiv:1612.01238}}\relax
\mciteBstWouldAddEndPuncttrue
\mciteSetBstMidEndSepPunct{\mcitedefaultmidpunct}
{\mcitedefaultendpunct}{\mcitedefaultseppunct}\relax
\EndOfBibitem
\end{mcitethebibliography}

\newpage

\end{document}